\documentstyle[preprint,prl,aps]{revtex}

\begin{document}
\draft

\title{Optical Theorem and the Inversion of Cross Section
Data for Atom Scattering from Defects on Surfaces
}

\author{D.A. Hamburger$^{a,b}$ and R.B. Gerber$^{b,c}$}
\address{
$^a$Department of Physics, The Hebrew University of
Jerusalem, Jerusalem
91904, Israel\\
$^b$The Fritz Haber Center for Molecular Dynamics, The
Hebrew University of
Jerusalem, Jerusalem 91904, Israel\\
$^c$Department of Chemistry, University of California -
Irvine, Irvine, CA
92717, USA
}

\maketitle

\begin{abstract}
\newline{}
The information content and properties of the cross section
for atom
scattering from a defect on a flat surface are
investigated. Using the
Sudden approximation, a simple expression is obtained that
relates the
cross section to the underlying atom/defect interaction
potential. An
approximate inversion formula is given, that determines the
shape
function of the defect from the scattering data.  Another
inversion
formula approximately determines the potential due to a
weak
corrugation in the case of substitutional disorder. An
Optical
Theorem, derived in the framework of the Sudden
approximation, plays a
central role in deriving the equations that conveniently
relate the
interaction potential to the cross section. Also essential
for the
result is the equivalence of the operational definition for
the cross
section for scattering by a defect, given by Poelsema and
Comsa, and
the formal definition from quantum scattering theory. This
equivalence
is established here. The inversion result is applied to
determine the
shape function of an Ag atom on Pt(111) from scattering
data.
\end{abstract}

\newpage

\section{Introduction}
The pioneering work by Poelsema, Comsa and
coworkers\cite{Comsa:d1,Comsa:d2,Comsa:d3,Comsa:d4,Comsa:d5}
on atom-surface
scattering in the last decade, has led to the introduction
and application of
the concept of the cross section of a defect on a surface
in studies of
questions related to surface disorder. Extensive
experimental work in the
field (as summarized in ref.\onlinecite{Comsa:book}) has
focused primarily on
the measurement of the dependence of the cross section on
the surface
parameters, usually the coverage dependence. Some efforts
have also been
directed to the dependence on the scattering parameters,
such as incidence
energy and angle. These cross section studies have
attracted significant
theoretical
interest,\cite{Jonsson2,Liu,Gum,Heller,Benny:3,Bos,Gum:1,Benny:cs}
concentrating almost exclusively on the dependence on the
scattering
parameters. Relatively little attention has been paid to
the inverse question
of the {\em {information content}} of the cross section.
Most work in this
direction has been experimental, again focusing on the use
of cross section
data to study {\em coverage-dependent} questions, such as
the onset of Ostwald
ripening,\cite{Comsa:Ostwald} and whether growth proceeds
in a layer-by-layer
fashion.\cite{Comsa:layers,Yang,Xu,Dastoor,Hinch,Safron}
With very few
exceptions,\cite{Benny:cs,me:Heptamers} most studies to
date do not deal with
the issue of the {\em structural} information contained in
the cross
section. For instance, can the cross section be used to
extract the shape of
an object on the surface, or its electron density profile?
Can the interaction
potential between an atom and a surface defect be obtained
from the cross
section data? In the present paper, we address these
questions theoretically,
by focusing on what can be learned from studying the cross
section of an {\em
individual} defect on a surface. \\
To make contact with experiment, through the operational
definition of the
cross section proposed by Poelsema and
Comsa,\cite{Comsa:book} it is required
to show the equivalence of the latter definition to the
standard formal one
from scattering theory, which we shall employ. Here we will
establish this
equivalence.  An analytical study of the cross section is
most easily
performed by relating the cross section through an optical
theorem to the
scattering amplitude. This has been done by several
authors.\cite{Xu,Jonsson2,Gum} We derive an optical theorem
for atom-surface
scattering by use of the Sudden Approximation
(SA),\cite{Benny:review1} which
has two significant advantages: it leads to a simple
explicit form for the
cross section, and is expected to be fairly accurate, as
the SA performs best
for the calculation of specular amplitudes. By employing
our optical theorem
result, we will answer a number of specific questions
related to the
information that can be extracted from the single defect
cross section.\\
First we shall examine the relative role of the long range
dispersion forces
{\sl vs} the short range repulsive part of the potential in
determining the
interaction of a He atom with an adsorbed defect. It is a
well-recognized fact
in surface scattering
theory,\cite{Jonsson1,Jonsson2,Benny:3,Comsa:book} that
the magnitude of the total cross section of a defect is
dominated by small
angle deflections caused by the long range forces. We will
show that they are
mostly effective in setting the magnitude of the cross
section, whereas at
least for the high-energy regime the short-range forces
tend to determine its
qualitative shape.\\
A second issue that will be pursued is whether the cross
section be used to
perform an inversion of the He-defect interaction
potential. An affirmative
answer will be given in two important cases: the cross
section can be used to
invert the semiclassical {\em shape function} of a defect
on a surface, and to
approximately invert the interaction potential of a He atom
with a weak
corrugation, e.g., due to dilute substitutional disorder.
The results are
applied to obtain the shape function of an Ag atom on
Pt(111), using cross
section data from P. Zeppenfeld.\cite{Zeppenfeld} \\
The structure of the article is as follows: we discuss the
standard and
operational definitions of the cross section in
section~\ref{definition}, and
prove their equivalence. In section~\ref{optical}, we
briefly review the SA,
and derive the optical theorem for atom-surface scattering
within the SA. We
then proceed to discuss a number of applications of the
optical theorem, in
section~\ref{applications} The first is a discussion of the
influence of the
short range repulsive part of the atom-surface interaction
potential. The
second is perhaps the central result of this paper: an
inversion scheme for
the shape function of an adsorbate on a surface from cross
section
measurements. We apply this result to data for Ag/Pt(111).
The third
application is the approximate inversion of the potential
from the cross
section data. Concluding remarks are presented in
section~\ref{summary}

\section{Definition of the Cross Section}
\label{definition}

Consider a scattering experiment of an atom beam with
wave-vector ${\bf
k}=(k_x,k_y,k_z)$, impinging on a flat surface (the $z=0$
plane). Let ${\bf
R}=(x,y)$ denote a vector in the plane. A typical
experimental setup will
involve a highly monochromatic beam with a lateral spread
in the surface
plane.  This beam is suitably described by a set of states
$|\Psi_{in}\rangle
= |\Phi_{\bf R}\rangle$. After impinging upon the surface,
the beam is
scattered, in part by the defect whose cross section we are
interested in. To
derive the cross section, we start from the definition of
the quantum
differential scattering cross section (QDCS) for two-body
collisions.\cite{Taylor} This definition generalizes quite
naturally for
atom-surface scattering, as we will demonstrate. Let
$w(d\Omega \leftarrow
\Phi_{\bf R})$ denote the probability that a particle
incident upon a target
as a wave packet $\Phi$, displaced laterally by ${\bf R}$
with respect to the
origin, emerges after an elastic collision in the solid
angle $d\Omega$. Then
the average number of observed scatterings into $d\Omega$
is:

\begin{eqnarray}
N_{sc}(d\Omega) = \sum_i w(d\Omega \leftarrow \Phi_{{\bf
R}_i}) \approx \int
d^2 R \: n_{inc} w(d\Omega \leftarrow \Phi_{\bf R})
\nonumber
\end{eqnarray}

\noindent where $n_{inc}$ is the incident density, which we
assume to be
uniform over an area $A$, i.e., $n_{inc} = N_{inc}/A$, so
that:

\begin{equation}
N_{sc}(d\Omega) = {N_{inc} \over A} {d\sigma \over d\Omega}
d\Omega
\label{eq:QDCS}
\end{equation}

\noindent where $d\sigma/d\Omega$ is the differential cross
section, defined
as:

\begin{equation}
{d\sigma \over d\Omega} d\Omega \equiv \sigma(d\Omega) =
\int_A d^2R \: w(d\Omega \leftarrow \Phi_{\bf R})
\label{eq:sigma-def}
\end{equation}

\noindent Thus, as is well known, the QDCS is seen to be a
measure of the
target area
effective in scattering the incident beam. It must be
emphasized that the
definition of the QDCS excludes the forward direction.
Classically speaking,
the reason is that one cannot distinguish a forwardly
scattered particle from
one that did not interact with the target at all. The
analogue in surface
scattering from a defect, is that one cannot distinguish a
{\em specularly}
scattered particle from one that was scattered by the flat
part of the
surface. More formally, the reason is that inclusion of the
forward direction
would lead to a divergence of the cross section. We would
now like to discuss
the connection between the latter definition, which is
widely accepted for
two-body processes, and the operational definition proposed
by Poelsema and
Comsa\cite{Comsa:book} for atom-surface scattering, and
applied extensively
for cross section calculations from diffraction data. Their
formula reads:

\begin{equation}
\Sigma_{op} = -\lim_{n \rightarrow 0} {1 \over I_0} {dI_s
\over dn}
\label{eq:Comsa}
\end{equation}

\noindent where $\Sigma_{op}$ is the {\em total} scattering
cross section,
$I_s$, $I_0$ represent the specular scattering intensity
from the corrugated
and flat
(smooth, defect-free) surfaces respectively, and $n$ is the
number of defects
per total surface area. Xu {\sl {et al.}}\cite{Heller}
employed this formula
to derive an expression for the cross section based on
Gaussian wave packets,
and showed that their expression can be interpreted as an
optical theorem. We
will show that the operational definition
(eq.(\ref{eq:Comsa})), is in fact
{\em identical} to the standard definition of quantum
scattering theory
(eq.(\ref{eq:sigma-def})), without recourse to a specific
method of
calculating the quantities that appear in the formulas.
This is true, provided
the proper conditions are established to relate two-body
and atom-surface
collisions. The conditions under which the QDCS is defined,
are:\cite{Taylor}

\begin{itemize}
\item{
the incident particles have a sharply defined momentum
${\bf p}_0$ and are
randomly displaced in a plane perpendicular to ${\bf p}_0$,
}
\item{
the distribution of scatterers must be such as to avoid
coherent scattering
off two or more centers.
}
\end{itemize}

When interpreted in terms of atom-surface scattering, we
see that the first
condition may be traded by random lateral displacements
parallel to the
surface plane (maintaining a sharply peaked momentum).  The
second requires
that the defects are distributed very dilutely. This is,
however, exactly the
assumption underlying the operational definition, and of
course holds for
scattering off a {\em single} defect. Thus the conditions
assumed for the
definition of the two-body cross section are naturally
extendible to the
atom-surface case, and suit the operational definition. In
order to
demonstrate the equivalence of the definitions, we first
replace
${N_{sc}(\Omega)/N_{inc}}$ by ${I(d\Omega)/I_0}\, d\Omega$
in
eq.(\ref{eq:QDCS}), where $I$ is the {\em off}-specular
intensity. $I_0$ is
the incident intensity, which is clearly equal to the total
scattered
intensity in the case of a {\em flat} surface, and is hence
indeed the same
quantity as defined in the discussion of the operational
definition above.
With these replacements, the total cross section is:

\begin{equation}
\Sigma \equiv \int {d\sigma \over d\Omega} \, d\Omega = {A
\over I_0} \int
I(\Omega) \, d\Omega
\label{sigma-total}
\end{equation}

\noindent Next consider taking the limit in the operational
definition. This
can be done
by assuming that only one defect is present on the surface,
so that for
all practical purposes $dn = {1/A}$, $A$ being the total
surface area,
and $dI = I_s - I_0$. Then from eq.(\ref{eq:Comsa}):

\begin{equation}
\Sigma_{op} = {A \over I_0} (I_0 - I_s)
\label{eq:Comsa-limit}
\end{equation}

\noindent But clearly, $I_0 - I_s = \int I(\Omega) \,
d\Omega$, so that
$\Sigma_{op} =
\Sigma$, as claimed.

\section{Optical Theorem for Surface Scattering in the
Sudden Approximation}
\label{optical}
We very briefly review the SA. For more details see
e.g. the review by Gerber.\cite{Benny:review1}

\subsection{The Sudden Approximation}
\label{sudden}

Basically, the SA requires that the momentum transfer in
parallel to the
surface be small compared with the momentum transfer normal
to the surface,
i.e.\cite{Benny:Sud1,Benny:review2}

\begin{equation}
{\left|{\bf q'}-{\bf q}\right|} \ll 2k_z
\label{eq:Sudcond}
\end{equation}

\noindent where $k_z$ is the incident wave number in the z
direction, ${\bf
q}$ is the
incident wave vector in parallel to the surface plane, and
${\bf q'}$ is any
intermediate or final wave vector in parallel to the
surface plane which plays
a significant role in the scattering process. The S-matrix
element for
scattering from ${\bf q}$ ${\bf q'}$ may then be shown to
be given by the
function:

\begin{equation}
\langle{\bf q}|S|{\bf q'}\rangle = {1 \over A} \int_{\rm
surf} d{\bf R} \:
e^{2i\eta({\bf R})} \, e^{i \Delta {\bf q} \cdot {\bf R}}
\label{eq:sud}
\end{equation}

\noindent where ${\bf R}=(x,y)$ and $\eta({\bf R})$ is the
scattering phase
shift
computed for fixed ${\bf R}$, given in the WKB
approximation by:

\begin{equation}
\eta({\bf R}) = \int _{{z_t}({\bf R} )}^{\infty} dz\:
\left[\left({{k_z}^2} -
{{2\,m\,V({\bf R} ,z)}\over {{\hbar}^2}} \right)^{1 \over
2} -{k_z} \right] -
{k_z}\,{z_t}({\bf R} )
\label{eq:eta}
\end{equation}

\noindent Here $z_t({\bf R})$ is the classical turning
point for the integrand
in
eq.(\ref{eq:eta}).  The normalization is chosen such that
the unit-cell $A$ is
the average area available for one adsorbate, i.e, is
determined by the
coverage. This choice is suitable for low coverages, when
the simultaneous
interaction of a He atom with two neighboring adsorbates is
small. Of course,
an adsorbate may ``spill over'' beyond its unit-cell due to
the long range of
the potential, and hence the integration in
eq.(\ref{eq:sud}) is over the
entire surface. Condition (\ref{eq:Sudcond}) for the
validity of the Sudden
approximation is expected to break down for systems of high
corrugation. For
instance, certain isolated adsorbates on an otherwise flat
surface can
represent, for realistic parameters, a very substantial
local
corrugation. Nevertheless, previous calculations have shown
that the Sudden
approximation reproduces rather well many features of the
scattering from
isolated adsorbates.\cite{Benny:Sud2} Features for which it
breaks down are,
e.g, intensity peaks due to double collision events, in
which the incoming
atom first hits the surface and then the adsorbate (or vice
versa), which are
a particularly sensitive manifestation of a strong
corrugation (we note that a
double-collision version of the Sudden approximation has
recently been
developed\cite{me:DC}). However, due to the assumption
(\ref{eq:Sudcond}), the
Sudden approximation is particularly useful for the
evaluation of specular
intensities at incidence angles close to the specular
direction, as confirmed
by comparison with numerically exact wave packet
calculations.\cite{Benny:3}
In the context of the present work, only the specular
intensities are required
(which are obtained by setting ${\bf q}' = {\bf q}$ in
eq.(\ref{eq:sud})),
which constitutes the most favorable condition for the
Sudden.

\subsection{Derivation of the Optical Theorem}
\label{derivation}
The ensuing discussion closely follows that of
Taylor\cite{Taylor} for
two-body scattering. We proceed to derive the optical
theorem for surface
scattering within the SA.\\ The probabilities
$w(d\Omega\leftarrow\phi_{\bf
R})$ in eq.(\ref{eq:sigma-def}) are:

\begin{equation}
w(d\Omega \leftarrow \Psi_{in}) = d\Omega \,
\int_{0}^{\infty} p^2 dp \:
|\Psi_{out}({\bf p})|^2 \: \: \: \: \: \: \: ({\bf p} =
\hbar {\bf k})
\label{eq:w}
\end{equation}

\noindent since $w(d^3p \leftarrow \Psi_{in}) = d^3p
\,|\Psi_{out}({\bf
p})|^2$ is the
probability that long after the collision the particle
incident as
$\Psi_{in}$, emerging as $\Psi_{out}$, has momentum in
$({\bf p}, {\bf p} +
d{\bf p})$. Only the direction of ${\bf p}$ is of interest,
as we are assuming
elastic scattering. The ``in'' and ``out'' states are
related by the S-matrix:
$|\Psi_{out}\rangle = S\,|\Psi_{in}\rangle $. In the
momentum representation:

\begin{equation}
\Psi_{out}({\bf p}) = \int d^3p' \: \langle{\bf p}|S|{\bf
p'}\rangle \,
\Psi_{in}({\bf p'})
\label{eq:S}
\end{equation}

\noindent The combination of
eqs.(\ref{eq:sigma-def},\ref{eq:w}-\ref{eq:S})
yields the
differential cross section. Following the discussion in the
preceding section,
it can be seen that there is nothing which essentially
restricts the use of
eqs.(\ref{eq:sigma-def},\ref{eq:w}-\ref{eq:S}) to two-body
scattering, and
with the restriction of scattering angles to the half-space
above the surface
($z>0$), they may as well be applied to atom-surface
scattering. We now
proceed to calculate the differential cross section within
the Sudden
approximation.  Let us denote ${\bf p} = \hbar ({\bf
q},k_z)$, where ${\bf q}$
is the wave-vector component in parallel to the surface,
and $\Delta {\bf q}
\equiv {\bf q'}-{\bf q}$. We require the familiar
``scattering-amplitude''
$f$, in terms of which the Optical Theorem will be
expressed. To find its form
within the SA, we first write down a general expression for
the S-matrix
element, in the case of scattering from a surface with a
defect. This
expression must be composed of two terms, reflecting
specular (the analogue of
forward scattering in the gas phase) and off-specular
scattering. The first
arises in the case of scattering from the flat part of the
surface, whereas
the second is due to scattering from the defect. Energy is
assumed to be
conserved, so the general expression for the S-matrix
element $\langle{\bf
p}|S|{\bf p'}\rangle$ contains an on-shell delta-function
of the energy $E_p
\equiv p^2/2m$:

\begin{equation}
\langle{\bf p}|S|{\bf p'}\rangle = \delta(p_z + p_z') \,
\delta(\hbar \Delta
{\bf q}) + {i \over {2\pi m \hbar}} \delta(E_p - E_{p'})
f({\bf p}
\leftarrow {\bf p'})
\label{eq:S-general}
\end{equation}

\noindent The identification of the scattering-amplitude by
comparing
eqs.(\ref{eq:sud}),(\ref{eq:S-general}) is now made as
follows. First, the
phase-shift $\eta$ is constant for scattering from a flat
surface, and may
well be chosen zero. In this case the SA yields
$\langle{\bf q}|S|{\bf
q'}\rangle = \delta(\Delta {\bf q})$ (eq.(\ref{eq:sud})
with $\eta({\bf R}) =
0$), in agreement with the general form for specular
scattering. Second, we
consider the remainder after subtraction of the specular
part from the RHS of
eq.(\ref{eq:sud}):

\begin{equation}
g({\bf q} \leftarrow {\bf q'}) = {2\pi \over {i A}}
\int_{\rm surf} d{\bf R} \:
\left(e^{2i\eta({\bf R})}-1 \right) \, e^{i \Delta {\bf q}
\cdot {\bf R}}
\label{eq:g}
\end{equation}

\noindent so that:

\begin{equation}
\langle{\bf q}|S|{\bf q'}\rangle = {1 \over A}
\delta({\Delta {\bf q}}) + {i
\over 2\pi} g({\bf q} \leftarrow {\bf q'})
\label{eq:S-intermediate}
\end{equation}

\noindent Energy conservation is implicitly assumed in the
SA expression
eq.(\ref{eq:sud}). Still, comparing
eq.(\ref{eq:S-intermediate}) and
eq.(\ref{eq:S-general}), $g$ is not quite the scattering
amplitude yet, since
it does not include the action of the S-matrix on the
wave-function part in
perpendicular to the surface. The SA
derivation\cite{Benny:Sud1} yields only
the action on the part parallel to the surface. However, we
can obtain the
corresponding factor by comparison to the well known
partial-wave expansion
result. If we assume a plane wave He-beam at normal
incidence to a perfectly
flat surface with a spherically symmetric He-adsorbate
interaction, the
partial-wave expansion conditions apply. The gas-phase
expression for the
scattering amplitude is then:\cite{Taylor}

\begin{equation}
f({\bf k} \leftarrow {\bf k'}) = {{2 \pi} \over {i k}}
\sum_{l,m} Y_l^m
(\hat{{\bf k}}) \left[ e^{2i \delta_l (E_p)}-1 \right]
Y_l^m (\hat{{\bf k}}')^*
\label{eq:fPW}
\end{equation}

\noindent where $Y_l^m (\hat{{\bf k}})$ is the spherical
harmonic whose
argument
$\hat{{\bf k}}$ denotes the polar angle $(\theta,\phi)$ of
$\hat{{\bf k}}$,
and $\delta_l (E_p)$ is the phase-shift in the angular
momentum basis $\{ |
E,l,m \rangle \}$. Comparing
eqs.(\ref{eq:g}),(\ref{eq:fPW}), we see that the
contribution of the component perpendicular to the surface
is a factor of
$1/k$, so that the SA scattering amplitude is:

\begin{equation}
f({\bf p} \leftarrow {\bf p'}) = {1 \over k} g({\bf q}
\leftarrow
{\bf q'})
\label{eq:f}
\end{equation}

\noindent We emphasize at this point our strategy for
deriving the optical
theorem: the
analogue to the gas-phase forward {\sl vs} non-forward
scattering, is the
flat-surface {\sl vs} defect scattering. Within the SA,
this distinction is
easily made by separating the vanishing part from the
non-zero part of the
phase-shift. \\ We are now ready to calculate the
differential cross section
within the SA.  Substituting eq.(\ref{eq:S-general}) into
eq.(\ref{eq:S})
yields:

\begin{equation}
\Psi_{out}({\bf p}) = \Psi_{in}({\bf q}, -k_z) + {i \over
{2\pi m \hbar}} \int
d^3p' \: \delta(E_p - E_{p'}) \, f({\bf p} \leftarrow {\bf
p'}) \,
\Psi_{in}({\bf p}')
\end{equation}

\noindent The first term is the wave scattered from the
flat part of the
surface, the
second by the defect. Next we restrict observation to
defect scattering only.
Using the lateral displacement assumption from
section~\ref{definition} we
then obtain:

\begin{equation}
\Psi_{out}({\bf p}) = {i \over {2\pi m \hbar}} \int d^3p'
\: \delta(E_p -
E_{p'}) \, f({\bf p} \leftarrow {\bf p'}) \, e^{-{i \over
\hbar} {\bf
R}\cdot{\bf p'}} \Phi({\bf p'})
\label{eq:psi-out}
\end{equation}

\noindent Substituting eq.(\ref{eq:psi-out}) into
eq.(\ref{eq:w}) and then into
(\ref{eq:sigma-def}):

\begin{eqnarray}
\sigma(d\Omega) = {d\Omega \over {2\pi m^2 \hbar^2}} \int_A
d^2R \:
\int_0^{\infty} p^2 \, dp \: \times \nonumber \\
\int d^3p' \: \delta(E_p - E_{p'}) \, f({\bf p} \leftarrow
{\bf p'}) \, e^{-{i
\over \hbar} {\bf R}\cdot{\bf p'}} \Phi({\bf p'}) \: \times
\nonumber \\
\int d^3p'' \: \delta(E_p - E_{p''}) \, f^*({\bf p}
\leftarrow {\bf p''}) \,
e^{{i \over \hbar} {\bf R}\cdot{\bf p''}} \Phi({\bf p''})
\label{eq:sigma-mess}
\end{eqnarray}

\noindent The ${\bf R}$ integration yields $(2\pi)^2 \,
\delta({\bf q}'' -
{\bf q}')$.
It is easily checked, using some well known properties of
the delta-function,
that consequently:

\begin{equation}
\int d^2 R \: e^{{i \over \hbar} {\bf R} \cdot ({\bf p}'' -
{\bf p}')} \,
\delta(E_p - E_{p'}) \, \delta(E_p - E_{p''}) = (2\pi)^2
{{m \, \hbar} \over
{k_z'}} \delta({\bf p}'' - {\bf p}') \, \delta(E_p -
E_{p'})
\end{equation}

\noindent Inserting this into eq.(\ref{eq:sigma-mess}) and
applying the
momentum
delta-function, we obtain:

\begin{equation}
\sigma(d\Omega) = {\hbar d\Omega \over m \hbar}
\int_0^{\infty} p^2
dp \: \int d^3 p' \: {1 \over k_z'} \delta(E_p - E_{p'}) \,
|g({\bf q}
\leftarrow {\bf q}')|^2 \, |\Phi({\bf p}')|^2 {1 \over
{\left(k'\right)^2}}
\end{equation}

\noindent Next we use $\delta(E_p - E_{p'}) = {m \over p}
\delta(p - p')$ to
do the
radial integration over $p$:

\begin{equation}
\sigma(d\Omega) = d\Omega \int d^3 p' \: {1 \over
{k_z'\, k'}} \, |g({\bf q} \leftarrow {\bf q}') \,
\Phi({\bf p}')|^2
\end{equation}

\noindent where $|{\bf p}| = |{\bf p}'|$. To simplify the
last integral and
obtain the
standard form, we recall that $\Phi$ represents the
incident wave-function (at
zero lateral displacement). The simplification is obtained
by the physically
plausible assumption that $\Phi$ is {\em {sharply peaked}}
about the incidence
momentum ${\bf p}_0$. More quantitatively, we assume that
${1 \over {k_z'\,
k'}} g({\bf q} \leftarrow {\bf q}')$ does not vary
appreciably in the region
of $\Phi$'s peak, so that it can be taken outside of the
integral sign, and be
replaced by its values at ${\bf k}_0$. Using further the
normalization
condition $\int d^3 p' \: |\Phi({\bf p}')|^2 = 1$, we
finally obtain the
familiar looking result for the differential cross section:

\begin{equation}
\sigma(d\Omega \leftarrow {\bf k}_0) =  d\Omega {1 \over
{{k_0}_z\, k_0}}
\, |g({\bf q} \leftarrow {\bf q}_0)|^2
\label{eq:diff-sigma}
\end{equation}

\noindent where from eq.(\ref{eq:g}):

\begin{equation}
g({\bf q} \leftarrow {\bf q}_0) = {4\pi \over A} \int d{\bf
R} \:
\sin\left(\eta({\bf R})\right) \, e^{i\eta({\bf R})} \,
e^{i ({\bf q}_0 - {\bf
q}) \cdot {\bf R}}
\label{eq:f-final}
\end{equation}

\noindent What we are after is the total cross section for
scattering by a
defect. We
next employ the differential cross section result to
express this in terms of
an optical theorem for surface scattering. To do so, we use
the manifest
unitarity of the SA expression, eq.(\ref{eq:sud}).
Separating once again the
specular and non-specular parts of the S-matrix, now in
operator form: $S = 1
+ R$, and applying the unitarity condition, we have: $R \,
+ \, R^{\dagger} =
- R \, R^{\dagger}$. Inserting a complete set and taking
matrix elements:

\begin{equation}
\langle {\bf p}' |R| {\bf p} \rangle \, + \, \langle {\bf
p} |R| {\bf p}'
\rangle^* = - \int d^3 p'' \langle {\bf p}'' |R| {\bf p}'
\rangle^* \, \langle
{\bf p}'' |R| {\bf p} \rangle
\label{eq:R}
\end{equation}

\noindent But it is clear from eq.(\ref{eq:S-general})
that:

\begin{eqnarray}
\langle {\bf p}' |R| {\bf p} \rangle = {i \over {2\pi \, m
\, p}} g({\bf q}
\leftarrow {\bf q}') \, \delta(E_p - E_{p'}) \nonumber
\end{eqnarray}

\noindent Inserting this into eq.(\ref{eq:R}) and factoring
out a common
$\delta$:

\begin{equation}
{1 \over p} \left( g({\bf q}' \leftarrow {\bf q}) - g({\bf
q} \leftarrow {\bf
q}')^* \right) = {i \over {2\pi \, m}} \int d^3 p'' \:
\delta(E_p - E_{p''})
\, g({\bf q}'' \leftarrow {\bf q}')^* \, g({\bf q}''
\leftarrow {\bf q}) {1
\over {\left(p''\right)^2}}
\label{eq:almost-optical}
\end{equation}

\noindent The optical theorem results by considering the
specular scattering
(${\bf q} =
{\bf q}'$) and integrating over ${\bf p}''$. In contrast to
two-body
scattering, in the atom-surface scattering case only the
space above the
surface is available, which can be taken care of by a
factor of ${1 \over 2}$.
Then, using $\int_0^{\infty} dp' \: h(p') \, \delta(E_p -
E_{p'}) = {m \over
p} h(p)$ and performing the radial integration:

\begin{eqnarray}
\int d^3 p'' \cdots = {1 \over 2} \int d\Omega_{{\bf p}''}
\: \int_0^{\infty}
p''^2 \,dp'' \: \delta(E_p - E_{p''}) \, g^*({\bf q}''
\leftarrow {\bf q}') \,
g({\bf q}'' \leftarrow {\bf q}) {1 \over p''^2} \nonumber
\\ = {m \over {2 \,
p}} \int d\Omega_{{\bf p}''} \, g^*({\bf q}'' \leftarrow
{\bf q}') \, g({\bf
q}'' \leftarrow {\bf q}) \nonumber
\end{eqnarray}

\noindent Using this and the specularity condition in
eq.(\ref{eq:almost-optical}), we obtain:

\begin{equation}
{2i \over p} {\rm Im}\left[g({\bf q} \leftarrow {\bf
q})\right] = {i \over
{2\pi}} {1 \over 2p} \int d\Omega_{{\bf p}'} \: |g({\bf q}'
\leftarrow {\bf
q})|^2
\label{eq:Img}
\end{equation}

\noindent Comparing the RHS of eq.(\ref{eq:Img}) with that
of
eq.(\ref{eq:diff-sigma}), we observe that it is just the
{\em total}
cross section for scattering at incidence momentum ${\bf
p}$, i.e.:

\begin{eqnarray}
\Sigma({\bf p}) \equiv \int d\Omega_{{\bf p}'} \:
\sigma(d\Omega_{{\bf p}'}
\leftarrow {\bf p}) \nonumber = {1 \over {k \, k_z}} \int
d\Omega_{{\bf
p}'} \: |g({\bf q}' \leftarrow {\bf q})|^2 \nonumber
\end{eqnarray}

\noindent whence finally, the Optical Theorem:
\begin{equation}
\Sigma({\bf k}) = {8\pi \over {k \, k_z}} {\rm Im}
\left[g({\bf q} \leftarrow
{\bf q})\right]
\end{equation}

\noindent From eq.(\ref{eq:f-final}) we
obtain an explicit form for the total cross section:

\begin{equation}
\Sigma({\bf k}) = {32\pi^2 \over {A \, k \, k_z}} \int_{\rm
surf} d{\bf R} \:
\sin^2 \eta({\bf R})
\label{eq:sigma-sud}
\end{equation}

\noindent Equation (\ref{eq:sigma-sud}) is the optical
theorem for surface
scattering
within the SA. As a final remark, we did not consider the
effect of Bragg
scattering from a periodical underlying surface. As shown
by Xu {\sl et
al.},\cite{Heller} this contribution should be counted
along with the specular
amplitude, and its incorporation in the present framework
should pose no
particular difficulty.

\section{Applications}
\label{applications}
We now illustrate a number of applications of our result.

\subsection{Solvable Model of a Single Adsorbate}
\label{model}
It has been known for long that the long-range attractive
interaction
is dominant in determining the He-adsorbate cross
section.\cite{Jonsson1,Benny:3,Comsa:book,Gum} We wish to
identify the
contribution of the short-range repulsive forces to the
cross section. For
this purpose, we will completely ignore the long-range
attractive forces and
consider a hard wall He-adsorbate interaction, where the
adsorbate is
described by a shape function $\xi(R)$. Then the potential
takes the form:

\begin{eqnarray}
V({\bf r}) = \left\{ \begin{array}{ll}
	0        & \mbox{: $z \geq \xi({\bf R})$} \\
	{\infty} & \mbox{: $z < \xi({\bf R})$}
	\end{array}
\right. \nonumber
\end{eqnarray}

\noindent From the expression for the phase-shift
eq.(\ref{eq:eta}), it is
clear that in this case:

\begin{equation}
\eta({\bf R}) = -k_z \, \xi({\bf R})
\label{eq:eta-hard}
\end{equation}

\noindent We now consider a fully solvable model of a
single adsorbate, which
although
highly artificial, will provide us with valuable insight
regarding the role of
the repulsive interaction in determining the general nature
of the cross
section. Let us assume a cylindrically symmetric shape
function of the form:

\begin{eqnarray}
\xi({\bf R}) = h \, \zeta\left(|{\bf R}|/l\right)
\:\:\:\:\:\:\:\:
\zeta(\alpha) = \left\{ \begin{array}{ll}
	0          & \mbox{: $\alpha > 1$} \\
	1-\alpha^n & \mbox{: $0 \leq \alpha \leq 1$}
	\end{array}
\right.
\end{eqnarray}

\noindent where $h$ is the height of the adsorbate above
the surface
(effectively the
strength of the coupling between the He and the adsorbate).
 $l = \gamma \, L$
($0 \leq \gamma \leq 1$) is a characteristic range of the
He-adsorbate
interaction. The unit-cell area (see
section~\ref{optical}\ref{sudden}) is
taken as $A = \pi L^2$. The linear extent of the entire
surface is denoted
$R_s$. We will demonstrate this choice of $\zeta(\alpha)$
to be analytically
solvable in terms of known functions for $n=1,2,4$. For
$n=1$ one obtains a
cone, for $n=2,4$ the shape is a {\em convexly} deformed
cone. It has a
physically unreasonable sharp edge for $\alpha = 1$, yet
the cases $n=2,4$
should be a crude, but reasonable model of the main
features of the adsorbate
shape function. In order to check the influence of the
sharp edge, we may
consider instead of $\zeta$ a {\em concave} shape function,
which trades the
edge with a sharp tip at $\alpha = 0$:

\begin{eqnarray}
\xi({\bf R}) = h \, \kappa\left(|{\bf R}|/l\right)
\:\:\:\:\:\:\:\:
\kappa(\alpha) = \left\{ \begin{array}{ll}
	0            & \mbox{: $\alpha > 1$} \\
	(1-\alpha)^n & \mbox{: $0 \leq \alpha \leq 1$}
	\end{array}
\right.
\end{eqnarray}

\noindent This model is solvable for $n=1,2$.\\
Using the symmetry, we have from the expression for the
cross section within
the SA (eq.(\ref{eq:sigma-sud})), for the convex case:

\begin{equation}
\Sigma({\bf k}) = {{64\pi^2 \, \gamma^2} \over {k \, k_z}}
\int_0^{\epsilon}
d\alpha \: \alpha \, \sin^2 \left[h \, k_z \, \zeta(\alpha)
\right]
\label{eq:sigma-single}
\end{equation}

\noindent where $\epsilon \equiv R_s/l$. Defining $\beta
\equiv h\,k_z$, and
letting
$I_n^{\zeta}(\beta)$, $I_n^{\kappa}(\beta)$ stand for the
integrals (without
the prefactors) in the convex and concave cases
respectively, they evaluate
to:

\begin{eqnarray}
I_1^{\zeta}(\beta) = I_1^{\kappa}(\beta) = {\epsilon^2
\over 4} +
{{\cos(2\beta)-\cos(2\beta(1-\epsilon))} \over {8 \,
\beta^2}} + {{\epsilon
\sin(2\beta(1-\epsilon))} \over {4 \, \beta}} \nonumber \\
I_2^{\zeta}(\beta) = {\epsilon^2 \over 4} -
{{\sin(2\beta)-\sin(2\beta(1-\epsilon^2))} \over {8 \,
\beta}} \nonumber \\
I_2^{\kappa}(\beta) = {\epsilon^2 \over 4} + {{1 \over 4}}
{{\sqrt{\pi \over
\beta}\,\left[ F_c\left(-2\,\sqrt{\beta/\pi}\right) -
F_c\left(2\,\sqrt{\beta/\pi} (\epsilon-1)\right)\right]}} +
{{\sin(2\,\beta) - \sin(2\,\beta (\epsilon-1)^2)}\over
{8\,\beta}} \nonumber \\
I_4^{\zeta}(\beta) = {\epsilon^2 \over 4} - {{1 \over 8}}
\sqrt{\pi \over
\beta}  \left( \cos(2\beta) \, F_c\left(2\epsilon^2\,
\sqrt{\beta/\pi}\right)
+ \sin(2\beta) \, F_s\left(2\epsilon^2\,
\sqrt{\beta/\pi}\right) \right)
\end{eqnarray}

\noindent where $F_c$, $F_s$ are the Fresnel integrals:

\begin{eqnarray}
F_c(x) = \int_0^x \cos \left({{\pi \, y^2} \over 2} \right)
\, dy
\:\:\:\:\:\:\:\ F_s(x) = \int_0^x \sin \left({{\pi \, y^2}
\over 2} \right) \,
dy \nonumber
\end{eqnarray}

\noindent An analysis of the models in the simple case of
$\epsilon=1$ yields
the
following results (see figure \ref{fig:1} - cone, figures
\ref{fig:2},
\ref{fig:3} - convex cases, figure \ref{fig:4} - concave
cases):

\begin{itemize}
\item{
In both the concave and convex cases the cross section
generally decreases
with increasing incidence wave-number $k_z$.
}
\item{\noindent
The convex model exhibits more noticeable oscillations.
These are due to
interference between He particles striking the top of the
adsorbate and the
surface. Since the top has a larger surface area in the
convex case, the
increased oscillations are expected.
}
\item{\noindent
The cross section is generally larger in the convex case.
This is also
expected due to the larger surface area of the top.
}
\item{\noindent
The cross section has a finite limit for $k_z \rightarrow
0$, which can easily
be found by a first-order Taylor expansion of the $\sin^2$
in
eq.(\ref{eq:sigma-single}). This limit is as a matter of
fact beyond the
region of validity of the SA. However, relying on the
impressive success of
the SA under unfavorable conditions of significant
corrugation,\cite{Benny:Sud2} we consider the limit anyway:
its dependence on
the corrugation parameter $h$ and the concavity/convexity
parameter $n$ is of
high interest. We find:

\begin{eqnarray}
\Sigma^{\zeta}(k \rightarrow 0) =
{{32{{\pi }^2}\,{h^2}\,{n^2}} \over {\left( 1 + n \right)
\,\left( 2 + n
\right) }} \:\:\:\:\:\:\:\:\: {\rm (convex)} \nonumber \\
\Sigma^{\kappa}(k \rightarrow 0) = {{32\pi^2\,{h^2}} \over
{\left( 1 + n
\right) \,\left( 1 + 2\,n \right) }} \:\:\:\:\:\:\:\:\:
{\rm (concave)}
\nonumber
\end{eqnarray}

\noindent The corrugation parameter $h$ is thus seen to be
of major importance
in
determining the cross section, which in the limit of low
incidence energy
scales as its square. The dependence on the exponent $n$
is, as expected,
opposite in the concave {\sl vs.} convex case (see figure
\ref{fig:5}): in the
former, the cross section monotonically decreases with $n$,
whereas it
increases in the latter.  This is once again due to size of
the surface area
of the adsorbate top in each case.
}
\item{
Due to the absence of the attractive part in our model, the
resulting cross
section values are too small. Typical values for e.g.
CO/Pt(111)\cite{Comsa:d2,Comsa:d3} are in the range
$100-300\AA^2$. This deficiency of the hard wall model was
already observed by
Jonsson {\sl {et al.}}\cite{Jonsson1,Jonsson2}
}
\item{
The integrals $I_n^{\zeta}(\beta)$, $I_n^{\kappa}(\beta)$
tend asymptotically
to a constant value with $k_z$. Therefore the decrease in
the cross section is
entirely (apart from oscillations) due to the $k_z^{-2}$
factor.
}
\end{itemize}

\noindent In conclusion, it appears that the cross sections
we obtained
exhibit all the
general features of the experimentally observed ones: the
oscillations, the
decrease with increasing incidence energy and the finite
limit for vanishing
incidence energy. We conclude that while the long-range
attractive part is
important in the setting the {\em magnitude} of the cross
section values (the
stronger the interaction, the larger the cross section), at
least for the
high-energy regime the energy dependent features are
already determined by the
short-range repulsive interaction. \\
In the next section we will significantly simplify the
relation
eq.(\ref{eq:sigma-single}) between the cross section and
the shape function,
without any further approximations, in order to be able to
eventually invert
the shape function from cross section measurements.

\subsection{Inversion of Shape Function from Measurement of
Cross Section}
\label{inversion}
In this section we will demonstrate how the optical theorem
result may be used
to approximately solve the inversion problem for the
He-adsorbate potential.
The SA was already employed by Gerber, Yinnon and
coworkers\cite{Benny:inversion1,Benny:inversion2} for an
approximate inversion
of the potential, by using the full angular intensity
distribution.  Here we
will demonstrate a less general result for the inversion of
the adsorbate
shape function, which will, however, have the advantage of
relying on a
simpler measurement, of the specular intensities alone. \\
We consider the
case of a cylindrically symmetric adsorbate. If our
unit-cell and the surface
are circles of radii $L$ and $R_s$ respectively, then from
eq.(\ref{eq:sigma-sud}) we have for the adsorbate cross
section:

\begin{equation}
\Sigma(k_z, \theta_{in}) = {32\pi^2 \over {\pi L^2 \, k_z^2
\cos(\theta_{in})}}
2\pi \int_0^{R_s}  \sin^2 \eta(R) \, R\, dR = {1 \over
{\alpha \,k_z^2}}
\left[ {1 \over 2} R_s^2 \,-\, \int_0^{R_s}  \cos\left(2
\eta(R)
\right) R \, dR \right]
\label{eq:sigma-invert}
\end{equation}

\noindent where

\begin{eqnarray}
\alpha \equiv {{L^2 \, \cos(\theta_{in})} \over {32\pi^2}}
\nonumber
\end{eqnarray}

\noindent (We note in passing that the often
discussed\cite{Jonsson2}
$\cos(\theta_{in})$ factor, $\theta_{in}$ being the angle
between the incident
direction and the surface normal, arises naturally in our
formalism). Our
demonstration in section~\ref{definition} that this cross
section is
equivalent to that measurable by a specular attenuation
experiment, ensures
that the last equation can be regarded as a simple integral
equation for the
phase-shift $\eta$ in terms of measurable quantities $R_s,
L$ and $\Sigma(k_z,
\theta_{in})$. The measurable part is:

\begin{equation}
M(k_z) \equiv {1 \over 2} {R_s}^2 \,-\, \alpha \, k_z^2\,
\Sigma(k_z, \theta_{in})
\label{eq:M}
\end{equation}

\noindent Let us now make the hard wall assumption again,
which leads to the
form of
eq.(\ref{eq:eta-hard}) for the phase-shift. By doing so, we
will be able to
obtain the form of the shape function $\xi({\bf R})$ of the
adsorbate. Of
course the concept of a shape function is rather
artificial, yet it is of
great interest in characterizing surface roughness, and is
not unrelated to
what is obtained in STM measurements. Defining for
convenience:

\begin{equation}
\Theta(R) = 2\, \xi(R) ,
\label{eq:theta}
\end{equation}

\noindent we can express the integral part as:

\begin{equation}
I_{\Theta} \equiv \int_0^{R_s} R \, dR \cos\left(k_z
\Theta(R) \right)
\label{eq:I}
\end{equation}

\noindent so that eq.(\ref{eq:sigma-invert}) becomes:
\begin{equation}
I_{\Theta} = M(k_z)
\label{eq:I-M}
\end{equation}

\noindent We next assume $\Theta(R)$ to be a monotone,
single-valued function,
so that
we can consider its inverse $R(\Theta)$ (see
figure~\ref{fig:6}). Thus:

\begin{equation}
I_{\Theta} = \int_{\Theta(0)}^{\Theta(R_s)} d\Theta \:
{dR(\Theta) \over
d\Theta} R(\Theta) \, \cos( k_z \Theta) = {1 \over
\sqrt{2\pi}}
\int_{\Theta(0)}^{\Theta(R_s)} d\Theta \: F(\Theta) \,
\left( e^{i \, k_z \,
\Theta} + e^{-i \, k_z \, \Theta} \right)
\label{eq:I-Theta}
\end{equation}

\noindent where we defined:

\begin{equation}
F(\Theta) = {\sqrt{2\pi} \over 4}
{d\left(R^2(\Theta)\right) \over d\Theta}
\label{eq:F}
\end{equation}

\noindent $R(\Theta)$ is confined, so it is convenient to
define a ``box''
function:

\begin{eqnarray}
b(\Theta) = \left\{ \begin{array}{ll}
	1        & \mbox{: $\Theta(R_s) \leq \Theta \leq
\Theta(0)$} \\
	0        & \mbox{: ${\rm else}$}
	\end{array}
\right.
\label{eq:b}
\end{eqnarray}

\noindent which allows us to write $I_{\Theta}$ as a
Fourier transform of a
product:

\begin{equation}
I_{\Theta} = {\cal F}\left[ F({\Theta}) \, b({\Theta}); \:
k_z\right] \:+\:
{\cal F}^{-1} \left[ F({\Theta}) \, b({\Theta}); \: k_z
\right]
\label{eq:I-FT}
\end{equation}

\noindent where we introduced a notation for the Fourier
transform,

\begin{eqnarray}
{\cal F}\left[f(x); \: y\right] = {1 \over \sqrt{2\pi}}
\int_{-\infty}^{\infty} dx\: f(x)\, e^{i \, x\, y}
\nonumber
\end{eqnarray}

\noindent To isolate $F(\Theta)$ we apply an inverse
Fourier transform to
eq.(\ref{eq:I-FT}), yielding:

\begin{equation}
F(\Theta) \, b(\Theta) \:+\: F(-\Theta) \, b(-\Theta) =
{\cal F}^{-1}
\left[ I_{\Theta}(k_z); \: \Theta \right]
\label{eq:I-IFT}
\end{equation}

\noindent But $\Theta > 0$ so that by definition of the box
function the
second term in
eq.(\ref{eq:I-IFT}) vanishes, and we obtain from it, upon
inserting the
definitions of $M$ (eq.(\ref{eq:M})) and $F$
(eq.(\ref{eq:F})):

\begin{equation}
{\sqrt{2\pi} \over 4} \, {d\left(R^2(\Theta)\right) \over
d\Theta} = \pi\,
{R_s}^2 \, \delta(\Theta) \:-\: \alpha\,{\cal F}^{-1}
\left[
k_z^2\,\Sigma(k_z,\theta_{in});\: \Theta \right]
\:\:\:\:\:\ {\rm for} \:\:\:
\Theta(R_s) \leq \Theta \leq \Theta(0)
\end{equation}

\noindent If we assume that the shape function never fully
vanishes
($\Theta(R_s) > 0$) then $\delta(\Theta) = 0$, so that
finally:

\begin{equation}
{d\left(R^2(\Theta)\right) \over d\Theta} = -{{L^2 \,
\cos(\theta_{in})} \over
{8\pi^2 \sqrt{2\pi}}} {\cal F}^{-1} \left[ k_z^2\,
\Sigma_{\theta_{in}}(k_z);
\: \Theta \right]
\label{eq:inversion}
\end{equation}

\noindent The last equation is the sought after inversion.
Its RHS should be
considered
as calculated from the experimental data on the cross
section as a function of
the incidence wavenumber, at a given incidence angle. The
inversion is
completed by solving for $R^2(\Theta)$ under the condition
$R(0) \rightarrow
\infty$, and retrieving the shape function $\xi(R) = {1
\over 2} \Theta(R)$.
\\
In practice, however, it is more reasonable to assume a
functional form for
$\Theta(R)$ with free parameters, to be fitted based on the
experimental data.
The somewhat artificial concept of a shape function may not
justify an attempt
to determine an actual {\em function} from the cross
section data, as implied
by eq.(\ref{eq:inversion}). This could be problematic due
to the discreteness
and finiteness of the available data set. If we assume this
approach, a
further simplification can be obtained by returning to
eq.(\ref{eq:I-Theta})
and integrating by parts (with $\Theta(R_s) = 0$ - an
excellent
approximation):

\begin{equation}
I_{\Theta} = {1 \over 2} \left[ R_s^2 - k_z
\int_0^{\Theta(0)} d\Theta\:
R^2(\Theta)\, \sin \left(k_z\, \Theta \right) \right]
\end{equation}

\noindent Equating this according to eq.(\ref{eq:I-M}) to
$M(k_z)$, and
assuming that we
have guessed a functional form for $\Theta(R)$ with a set
of free parameters
$\{ \alpha_i \}$, the inversion result may now be stated in
the form:

\begin{equation}
{{\cos(\theta_{in})} \over {(4 \pi)^2}}
\Sigma_{\theta_{in}}(k_z) =
g\left( k_z; \{ \alpha_i \} \right) \:\:\:\:\:\:\:\ g\left(
k_z; \{ \alpha_i
\} \right) \equiv {1 \over{L^2\,k_z}} \int_0^{\Theta(0)}
d\Theta\:
R^2(\Theta; \{ \alpha_i \})\, \sin \left(k_z\, \Theta
\right)
\label{eq:fit}
\end{equation}

\noindent The remaining task is to find the set $\{
\alpha_i \}$ which gives
the best fit
for the chosen form of $g$ to the cross section data. In
the next section we
shall implement this approach on experimental data.
Alternatively, the last
result may be viewed as a theoretical expression for the
cross section, within
the SA and a hard wall model.

\subsection{Fit of Shape Function for Ag on Pt(111)}
\label{Ag-fit}
We now employ our inversion result to obtain a fit for a
shape function of an
Ag atom on a Pt(111) surface, based on He scattering
measurements by P.
Zeppenfeld.\cite{Zeppenfeld} The shape function is a
legitimate and highly
relevant object of study in the field of surface roughness.
It basically
corresponds to an equipotential surface, and therefore
within a certain
approximation,\cite{Norskov1,Norskov2} also to the core
electron densities. As
a result, it is a complementary quantity to what is
measured in STM
experiments, where one probes the electron densities at the
Fermi
level. Assuming cylindrical symmetry, the shape function
should have the
general form depicted in figure~\ref{fig:6}: it should be
flat and smooth at
its top, and decay to zero height far away from the
nucleus. This leads us to
consider two simple forms for the shape function: a
Lorentzian and a Gaussian,
with two free parameters $l$, $\Theta_0$ each.  Thus:

\begin{equation}
\Theta_L(R) = {\Theta_0 \over {1+\left({R \over
l}\right)^2}} \:\:\:\:
{\rm or} \:\:\:\: \Theta_G(R) = \Theta_0 \, e^{-\left({R
\over l}\right)^2}
\label{eq:L-G}
\end{equation}

\noindent Inverting these for $R^2(\Theta)$ leads to
integrals $g\left( k_z;
\{ \alpha_i \} \right)$ (eq.(\ref{eq:fit})) expressible in
terms of known
functions:

\begin{equation}
g_L \left( k_z; \{l, \Theta_0 \}\right) = {l^2 \over L^2}
\left(
{{\left(\cos (k_z\, \Theta_0) -1 \right) } \over {k_z^2}} +
{{\Theta_0\, {\rm
si}(k_z\,\Theta_0)} \over k_z} \right)
\label{eq:gL}
\end{equation}

\begin{equation}
g_G \left( k_z; \{l, \Theta_0 \}\right) = {l^2 \over L^2}
{1 \over k_z^2}
\left( \Gamma + {\rm ln}(k_z\, \Theta_0) - {\rm ci}(k_z\,
\Theta_0) \right)
\label{eq:gG}
\end{equation}

\noindent where ${\rm si(x}), {\rm ci(x)}$ are the sine and
cosine integrals
respectively, and $\Gamma$ is Euler's constant (0.5772..).
Both $g_L$ and
$g_G$ contain an oscillatory part, expressing the
interference between parts
of the He beam striking the top of the adsorbate and the
flat surface, and
both decay as $k_z^{-2}$ for large $k_z$. The parameters
$l$ (the He-adsorbate
interaction range) and $L$ (the unit-cell extent) appear
only in the
combination $l^2/L^2$, and may hence be treated as a single
dimensionless
parameter $\gamma \equiv l/L$.  This result is easily seen
to hold for any
shape function of the form $\Theta \left( R/l \right)$, as
was already
demonstrated in section~\ref{applications}\ref{model} The
dependence of the
cross section on the He-adsorbate interaction range thus
contains no
surprises: the cross section scales as the square of this
range. It is the
adsorbate height $\Theta_0$, i.e., the coupling coefficient
between the He and
the adsorbate, which plays the interesting role. It
determines the frequency
of oscillations of the cross section, and affects its
magnitude.\\
Using the results for the Lorentzian and Gaussian models,
eqs.(\ref{eq:gL},\ref{eq:gG}) respectively, in the
expression for the cross
section, eq.(\ref{eq:fit}) we calculated best-fits for the
parameters
$\Theta_0$ and $\gamma$. The fits are shown along with the
empirical cross
section values\cite{Zeppenfeld} in figure~\ref{fig:7}. The
deviations are
1.47\% for the Lorentzian, 1.14\% for the Gaussian. The
fits do not follow the
oscillations in the data closely, but these are uncertain
anyway due to the
large experimental error ($ \sim 20\% $). The values found
are:\\
Lorentzian: $\Theta_0 = 1.51\AA$, $\gamma = 0.93$\\
Gaussian:   $\Theta_0 = 1.05\AA$, $\gamma = 1.83$\\
The values for the adsorbate ``height'' $\Theta_0$ are in
reasonable agreement
with those obtained from STM measurements.\cite{Zeppenfeld}
That $\gamma > 1$
for the Gaussian model expresses the fact that the
He-adsorbate interaction
range extends beyond the adsorbate unit-cell. In any
inversion problem, the
question of stability is major importance. In the present
case, convergence to
the fitted values is overall better for the Gaussian model,
which also has a
smaller deviation: the Gaussian model converges to the same
values for initial
guesses of $0<\Theta_0<5\AA$, whereas convergence is
obtained for the
Lorentzian model only for the smaller range of
$0<\Theta_0<3\AA$. The initial
value of the parameter $\gamma$ has almost no influence on
the
convergence. When compared over a large range of $k_z$
values (see
figure~\ref{fig:8}), the two models are seen to essentially
differ in their
prediction of the cross section merely by a shift. The
situation is different,
however, when we compare the sought-after quantity: the
shape function. The
corresponding shape functions are shown in
figure~\ref{fig:9}. The Lorentzian
model is much more peaked and narrow. Intuitively, it seems
that the Gaussian
shape is better suited to describe the adatom shape
function, in agreement
with the relative deviations and convergences.

\subsection{Inversion of Potential for Weakly Corrugated
Surfaces}
\label{inversion-weak}
Consider a surface of substrate type A, in which one of the
atoms is replaced
by an atom of type B, such that the radii differ only
slightly. This is an
example of a surface with weak corrugation due to
substitutional disorder. Let
the surface be flat and located at $z=0$. The
interaction potential with an incident He atom may be
written approximately as:

\begin{equation}
U(R,z) = U_s(z) + U_d(R^2+z^2) \, , \:\:\:\:\:\:\:\: |U_s|
\gg |U_d|
\label{eq:U}
\end{equation}

\noindent where ${\bf R}=(x,y)$, $V_s = {\hbar^2 \over
{2m}} U_s$ is the
contribution of
the surface, and $V_d = {\hbar^2 \over {2m}} U_d$ is the
interaction with the
defect, assumed spherically symmetric. We next assume
$V_s(z)$ is known and
wish to determine $V_d$ from the scattering data. We note
that a similar
inversion has already been considered by Gerber, Yinnon and
coworkers,\cite{Benny:inversion1,Benny:inversion2} but for
defectless,
crystalline surfaces. When the condition $|U_s| \gg |U_d|$
is used in
eq.(\ref{eq:eta}), a first order expansion yields:

\begin{eqnarray}
\eta(R) \approx \eta_0(R) + {1 \over 2} \eta_1(R) \nonumber
\\
\eta_0(R) = \int _{{z_t}(R)}^{\infty} dz\:
\left[\left({{k_z}^2} - U_s(z)
\right)^{1 \over 2} -{k_z} \right] - {k_z}\,{z_t}({\bf R} )
\nonumber \\
\eta_1(R) = -\int _{{z_t}(R)}^{\infty} dz\: {{U_d(R^2+z^2)}
\over {\left(
{{k_z}^2} - U_s(z) \right)^{1 \over 2}}}
\label{eq:eta01}
\end{eqnarray}

\noindent The turning-point function $z_t(R)$ may either be
measured by the
inversion
procedure described in
section~\ref{applications}\ref{inversion} for the shape
function $\xi$, or approximated as the solution to $V_s(z)
= E$, i.e.,
neglecting $V_d$ altogether. In either case $\eta_0(R)$ is
a known
quantity. Let us next use our approximation for $\eta$ in
eq.(\ref{eq:sigma-invert}). Then after a first order
expansion about
$2\eta_0$:

\begin{equation}
\alpha k_z^2 \, \Sigma(k_z,\theta_{in}) - {1 \over 2} R_s^2
+ \int_0^{R_s}
dR\: R\, \cos(2\eta_0) \approx \int_0^{R_s} dR\: R\,
\sin(2\eta_0(R)) \,
\eta_1(R)
\label{eq:1order}
\end{equation}

\noindent The LHS consists of measurable/known quantities
and we denote it
$M_1(k_z)$.
The RHS contains the unknown function $U_d$ which we are
after; we denote it
$I(k_z)$. Using the definition of $\eta_1$
(eq.(\ref{eq:eta01})):

\begin{equation}
I(k_z) = -\int_0^{R_s} dR\: R\, \sin(2\eta_0(R)) \: \int
_{{z_t}(R)}^{\infty}
dz\: {{U_d(R^2+z^2)} \over {\left( {{k_z}^2} - U_s(z)
\right)^{1 \over 2}}}
\label{eq:I-weak}
\end{equation}

\noindent A practical approach at this point would be to
expand $U_d(R^2+z^2)$
in some
convenient basis $\{ f_n(R^2+z^2) \}$: $U_d(R^2+z^2) = \sum
c_n f_n(R^2+z^2)$,
so that:

\begin{equation}
\sum c_n \int_0^{R_s} dR\: R\, \sin(2\eta_0(R)) \: \int
_{{z_t}(R)}^{\infty}
dz\: {{f_n(R^2+z^2)} \over {\left( {{k_z}^2} - U_s(z)
\right)^{1 \over 2}}} =
-M_1(k_z)
\end{equation}

\noindent This linear system can then be solved for as many
$c_n$ as there are
data
points (in $k_z$). A convenient choice of (non-orthogonal)
basis functions are
Gaussians, since then the double integral decouples into a
product (neglecting
the $R$-dependence of $z_t$). This approach should give a
reasonable
approximation to the defect potential. Powerful methods to
deal with related
numerical inversion problems, employing functional
sensitivity analysis, have
been developed by Ho and
Rabitz.\cite{Rabitz:inversion1,Rabitz:inversion2}\\
However, with some further assumptions it is also possible
to perform an
actual inversion for $U_d$, by combining the technique of
section~\ref{applications}\ref{inversion} and an Abel
transform. We proceed to
show this. \\
Our first two additional assumptions are:
\begin{itemize}
\item{The interaction with the surface is purely repulsive,
so that $\zeta =
U_s(z)$ is a monotonously decreasing function.}
\item{Also $\rho = \eta_0(R)$ is a monotone function. That
this is
reasonable can be seen by considering a hard-wall system;
then a monotonously
decreasing shape function (see
section~\ref{applications}\ref{inversion})
induces a monotonously increasing phase shift, by
eq.(\ref{eq:eta-hard}).}
\end{itemize}

Under these assumptions we may consider the inverse
functions $z(\zeta)$ and
$R(\rho)$. The transformation to the variable $\zeta$
decouples the
integration limits if we make the fair approximation that
$k_z^2 =
U_s(z_t(R))$ (i.e., neglecting $U_d$), for then:

\begin{eqnarray}
\int _{{z_t}(R)}^{\infty} dz\: {{U_d(R^2+z^2)} \over
{\left( {{k_z}^2} -
U_s(z) \right)^{1 \over 2}}} = \int_{k_z^2}^0 d\zeta \: {dz
\over {d\zeta}}
{{U_d(R,\zeta)} \over {\left( {{k_z}^2} - \zeta \right)^{{1
\over 2}}}}
\nonumber
\end{eqnarray}

\noindent and hence:

\begin{eqnarray}
I(k_z) = \int_{k_z^2}^0 d\zeta \: {\left( {{k_z}^2} - \zeta
\right)^{-{1 \over
2}}} f(\zeta) \:\:\:\:\:\:\:\:\:\ {\rm where} \nonumber \\
f(\zeta) = {dz \over {d\zeta}} \, \int_0^{R_s} dR\: R\,
\sin(2\eta_0(R)) \,
U_d(R,\zeta)
\label{eq:I-Abel}
\end{eqnarray}

\noindent The above expression has the form of the familiar
Abel transform,
for which an
explicit inversion exists:

\begin{equation}
f(\zeta) = -{1 \over \pi} \int_0^{\zeta} dk_z^2 \: {dI(k_z)
\over dk_z^2}
\left(\zeta-k_z^2\right)^{-{1 \over 2}}
\label{eq:f-inverse}
\end{equation}

\noindent We rewrite this, using $I = M_1$
(eq.(\ref{eq:1order})) and changing
variables from $R$ to $\rho$, as:

\begin{equation}
J \equiv -\int_{\rho(0)}^{\rho(R_s)} d\rho \:
U_d(\rho,\zeta)\, {dR \over
d\rho} \, R(\rho) \, \sin(2\rho) = G(\zeta,k_z)
\label{eq:J}
\end{equation}

\noindent where:

\begin{equation}
G(\zeta,k_z) = {1 \over \pi} {d\zeta \over dz}
\int_0^{\zeta} dk_z^2 \:
{dM_1(k_z) \over dk_z^2} \left(\zeta-k_z^2\right)^{-{1
\over 2}}
\label{eq:G}
\end{equation}

\noindent The function $G(\zeta,k_z)$ consists entirely of
known
quantities. We are
still left with the task of isolating the potential $U_d$.
But it is clear
that $J$ in eq.(\ref{eq:J}) is almost identical to
$I_{\Theta}$ in
eq.(\ref{eq:I-Theta}). In order to be able to use the
Fourier transform
technique applied to $I_{\Theta}$, we must now introduce
one additional
assumption:
\begin{itemize}
\item{$\eta_0(R)$ may reasonably well be approximated in
the ``hard-wall''
form: $\eta_0(R) = -k_z \xi(R)$. Here $\xi(R)$ is to be
considered known from
the inversion of cross section data as described in
section~\ref{applications}\ref{inversion} Hamburger {\sl et
al.}\cite{me:Heptamers} have shown that this form holds
quite well also for
realistic potentials.}
\end{itemize}

Changing variables to $\Theta = 2 \xi(R)$:

\begin{eqnarray}
J = \int_{\Theta(0)}^{\Theta(R_s)} d\Theta \: \left[
U_d(\Theta,\zeta) {dR
\over d\Theta} R(\Theta) \right] \sin(k_z \Theta) \nonumber
\\
= {1 \over \sqrt{2\pi}} \int_{\Theta(0)}^{\Theta(R_s)}
d\Theta \: F(\Theta) \,
\left( e^{i k_z \Theta} - e^{-i k_z \Theta} \right) \equiv
I_{\Theta}
\end{eqnarray}

\noindent where now:
\begin{equation}
F(\Theta) = {\sqrt{2\pi} \over {4 i}} {d(R^2(\Theta)) \over
d\Theta}
U_d(\Theta,\zeta)
\label{eq:F-weak}
\end{equation}

\noindent Repeating the arguments leading to
eq.(\ref{eq:inversion}) and
remembering that now $I_{\theta} = G(\zeta,k_z)$, we
finally obtain:

\begin{equation}
U_d(\Theta,\zeta) = {1 \over {{d(R^2(\Theta)) \over
d\Theta}}} {{4 i} \over
\sqrt{2\pi}} {\cal F}^{-1} \left[ G(\zeta,k_z); \Theta
\right]
\label{eq:U-inverted}
\end{equation}

\noindent Formally, this completes the inversion.

\section{Concluding Remarks}
\label{summary}
In this paper we investigated theoretically the cross
section of a defect on a
surface. We found that this quantity, when available from
experiments,
contains important information on the defect and its
interaction with an
incoming atom. We demonstrated that the cross section can
be used for such
purposes as an inversion to yield the shape function of a
defect, an
approximate inversion to give the interaction potential
with a substituted
atom, and evaluation of the role of the long-range
repulsive forces in the
interaction with an incident atom. The single most
important advantage of the
cross section is probably that it is so easily measurable,
by employing the
definition of the cross section proposed by Poelsema and
Comsa.\cite{Comsa:d1}
This operational definition was shown here to be equivalent
to the standard
formal definition of quantum scattering theory, and hence
either can be used
interchangeably. Using the formal definition and the Sudden
approximation,\cite{Benny:review1} we derived a form of the
Optical theorem
which is quite amenable to analytical study. Future work
may benefit from this
expression for the cross section. For example, it seems
that extracting
various probability distributions characterizing dilute
surface disorder (such
as a distribution of radii for hemispherical defects)
should prove possible
within this framework. Traditional approaches tended to
rely on differential
cross section measurements for inversion applications. Our
general message for
future work is that the total cross section, which is much
easier to obtain
experimentally, will be a highly fruitful subject of study
for the purpose of
both structural and dynamical characterization of isolated
defects. Indeed, a
very promising possibility for future work should be one of
measuring cross
sections for surface defects of interest, and using this
data in a
corresponding theoretical effort to extract information on
the position of the
defect, its geometric shape, and its interaction potential
with the incident
atom. The latter is especially important, since the
interaction potential
contains information on the electronic density structure of
the defect. We
emphasize that cross sections measurements, when performed
over a wide range
of energies and incidence angles, may well become a new
type of surface
microscopy. We are currently pursuing efforts along this
line, in cooperation
with the experiments of P. Zeppenfeld.

\acknowledgements
This research was supported by the German-Israeli
Foundation for Scientific
Research (G.I.F.), under grant number I-215-006.5/91 (to
R.B.G.). Part of this
work was carried out with support from the Institute of
Surface and Interface
Science (ISIS) at UC Irvine. The Fritz Haber Center at the
Hebrew University
is supported by the Minerva Gesellschaft f\"{u}r die
Forschung, Munich,
Germany. We are very grateful to Drs. P. Zeppenfeld and M.
A. Krzyzowski for
helpful discussions and allowing us to use their Ag/Pt(111)
data prior to
publication. We would also like to thank Dr. A.T. Yinnon
for very useful
discussions.

\begin{figure}
\caption{Cross section in the cone case ($n=1$), for two
values of the corrugation parameter $h$. The cross sections
are almost
identical, except for the behavior at small $k_z$.}
\label{fig:1}
\end{figure}

\begin{figure}
\caption{Cross section in the convex case ($n=2$), for two
values of the corrugation parameter $h$. The stronger the
corrugation, the
faster and more pronounced are the oscillations, and the
larger the cross
section for small $k_z$.}
\label{fig:2}
\end{figure}

\begin{figure}
\caption{Cross section in the convex case, for three
values of the convexity parameter $n$, at a constant
corrugation $h=1\AA$. The
higher the convexity, the faster and more pronounced are
the oscillations, and
the larger the cross section for small $k_z$.}
\label{fig:3}
\end{figure}

\begin{figure}
\caption{Cross section in the concave case for two
values of the concavity parameter $n$, at a constant
corrugation $h=1\AA$. The
larger the concavity, the smaller the cross section.
Oscillations are
virtually unnoticeable.}
\label{fig:4}
\end{figure}

\begin{figure}
\caption{Cross section for $k_z \rightarrow 0$, in the
concave
and convex cases, as a function of the concavity/convexity
parameter. The more
concave (i.e. peaked), the smaller the cross section. The
opposite happens the
more convex (i.e. rectangular) the shape function.}
\label{fig:5}
\end{figure}

\begin{figure}
\caption{General form of the shape function $\Theta(R)$ and
its inverse $R(\Theta)$. The shape function should have a
smooth top and
monotonously decrease over a substantial range. We ignore
the possibility of a
subsequent increase.}
\label{fig:6}
\end{figure}

\begin{figure}
\caption{Cross section data for single Ag atom on
Pt(111) (circles), along with fits for Lorentzian (solid
line) and Gaussian
(dashed line) shape function models. The agreement with the
Gaussian model is
somewhat better.}
\label{fig:7}
\end{figure}

\begin{figure}
\caption{Behavior of the cross sections of the Lorentzian
and
Gaussian models over a large range of incidence wave
numbers. The two models
predict a similar behavior, differing essentially by a
shift along the
y-axis.}
\label{fig:8}
\end{figure}

\begin{figure}
\caption{Shape functions of the Ag adatom for the
Lorentzian
and Gaussian models, using the parameters obtained by
fitting the predicted
cross sections to the measured one. The Gaussian model has
a smoother top and
is less narrowly peaked, in agreement with intuitive
expectation for the
``correct'' shape. Note the different scales along the x
and y-axes.}
\label{fig:9}
\end{figure}

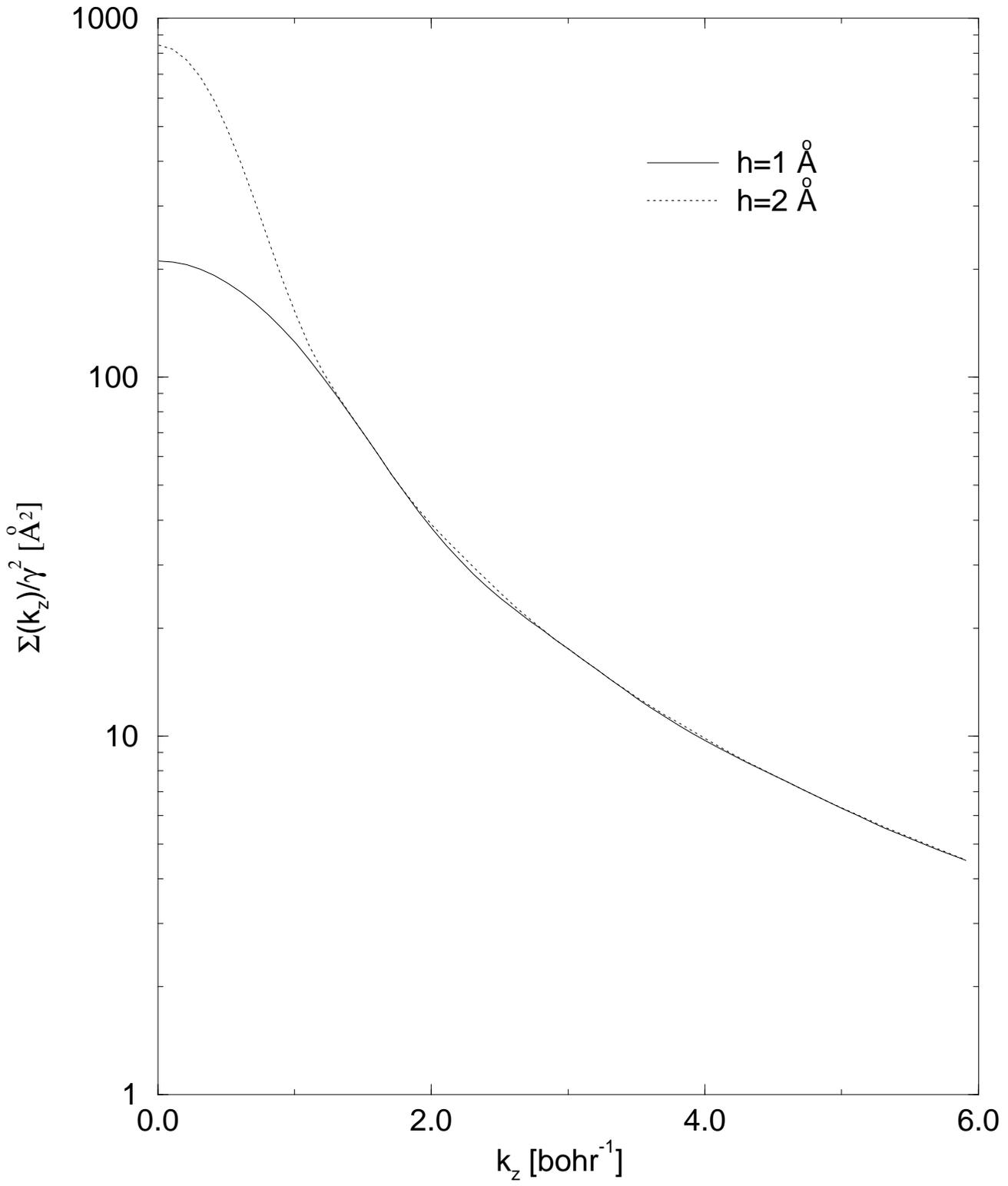
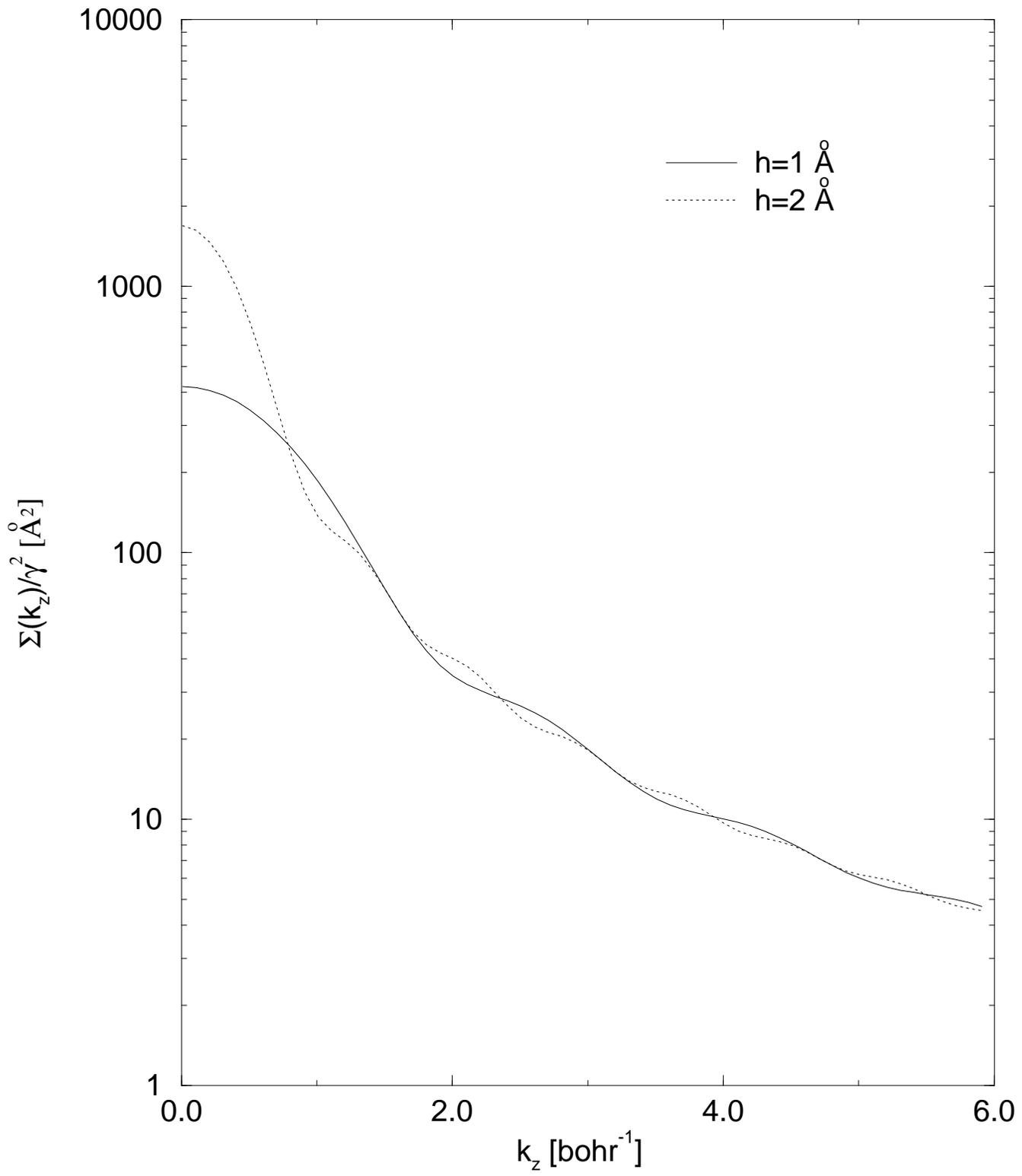
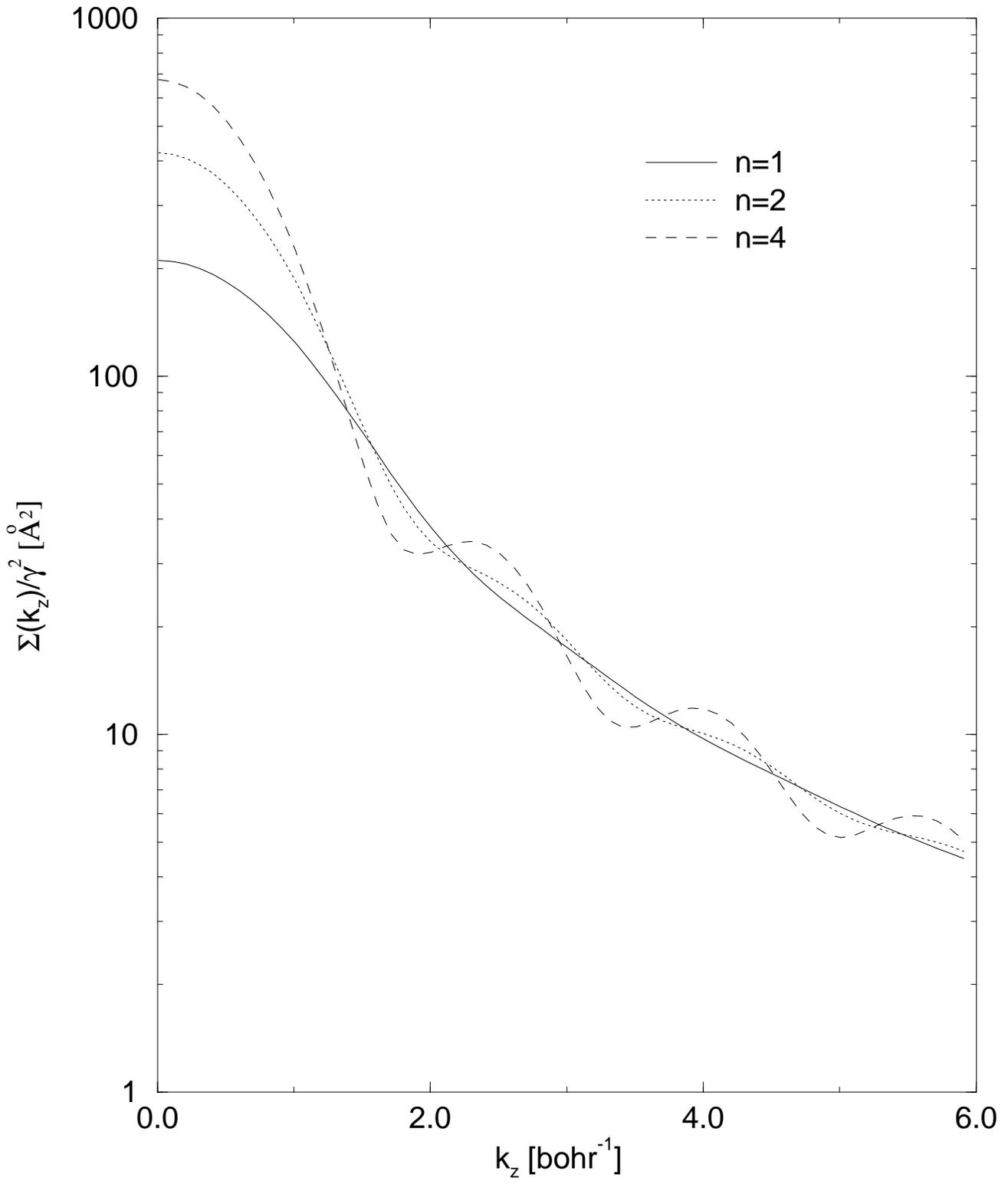
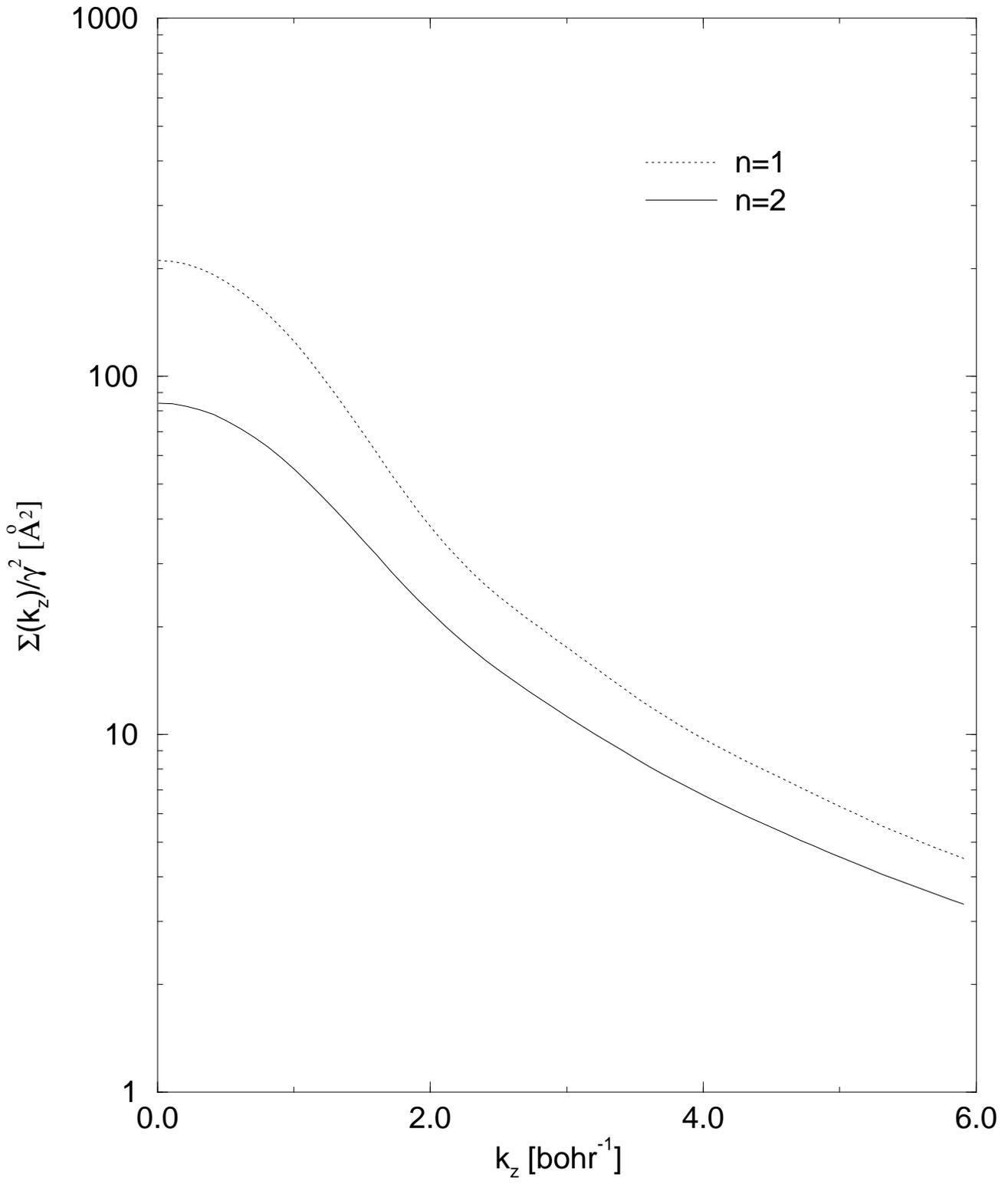
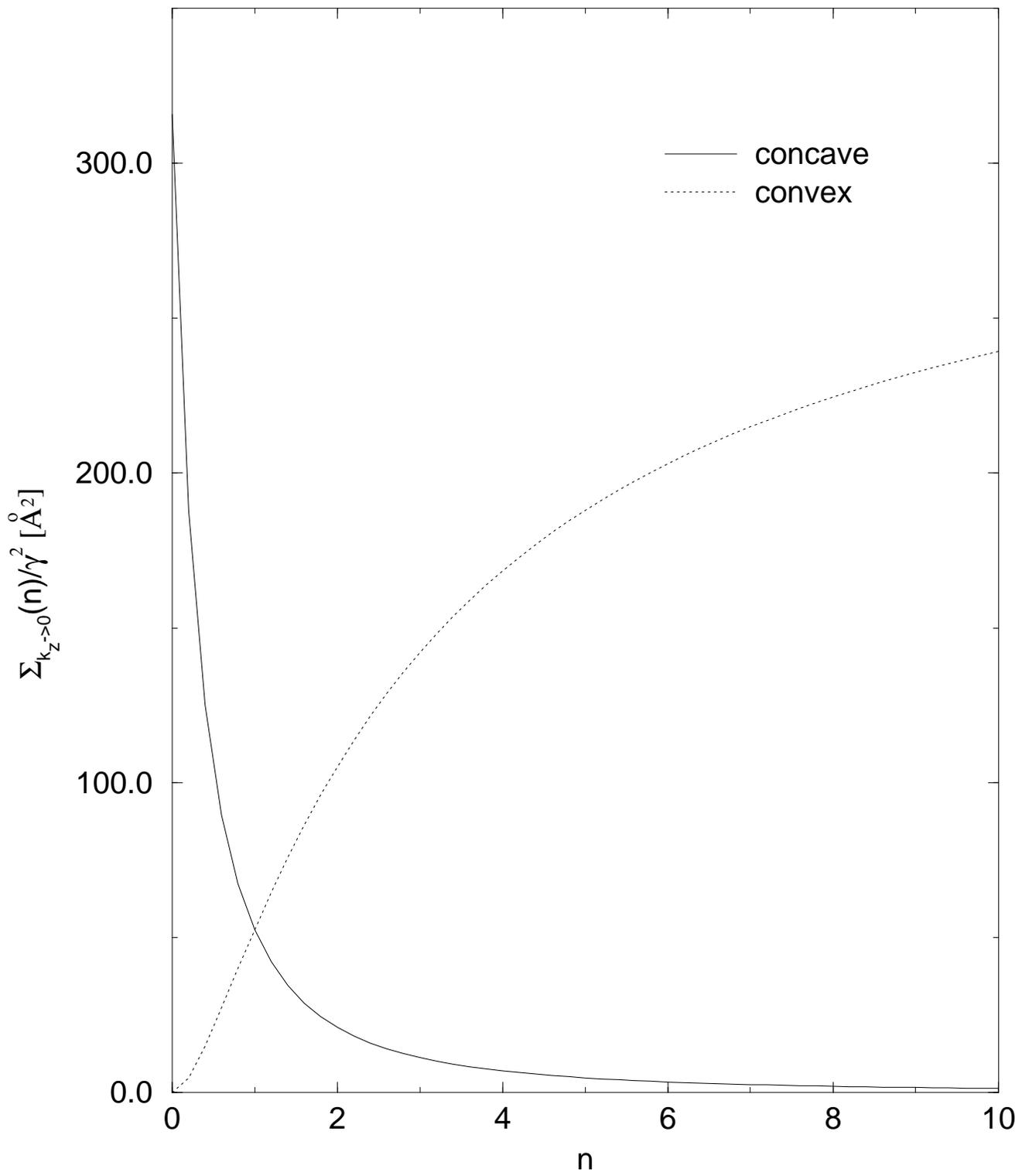
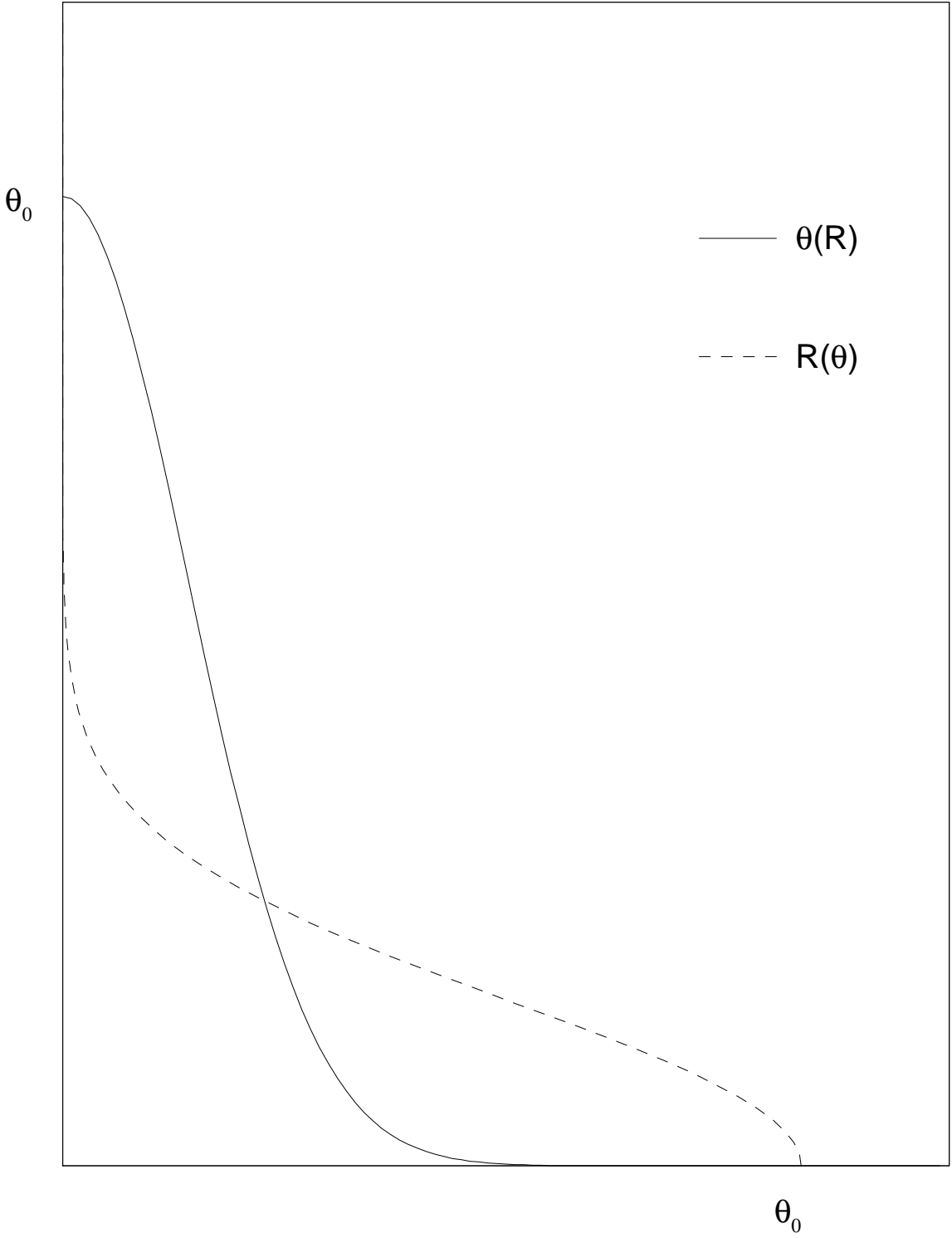
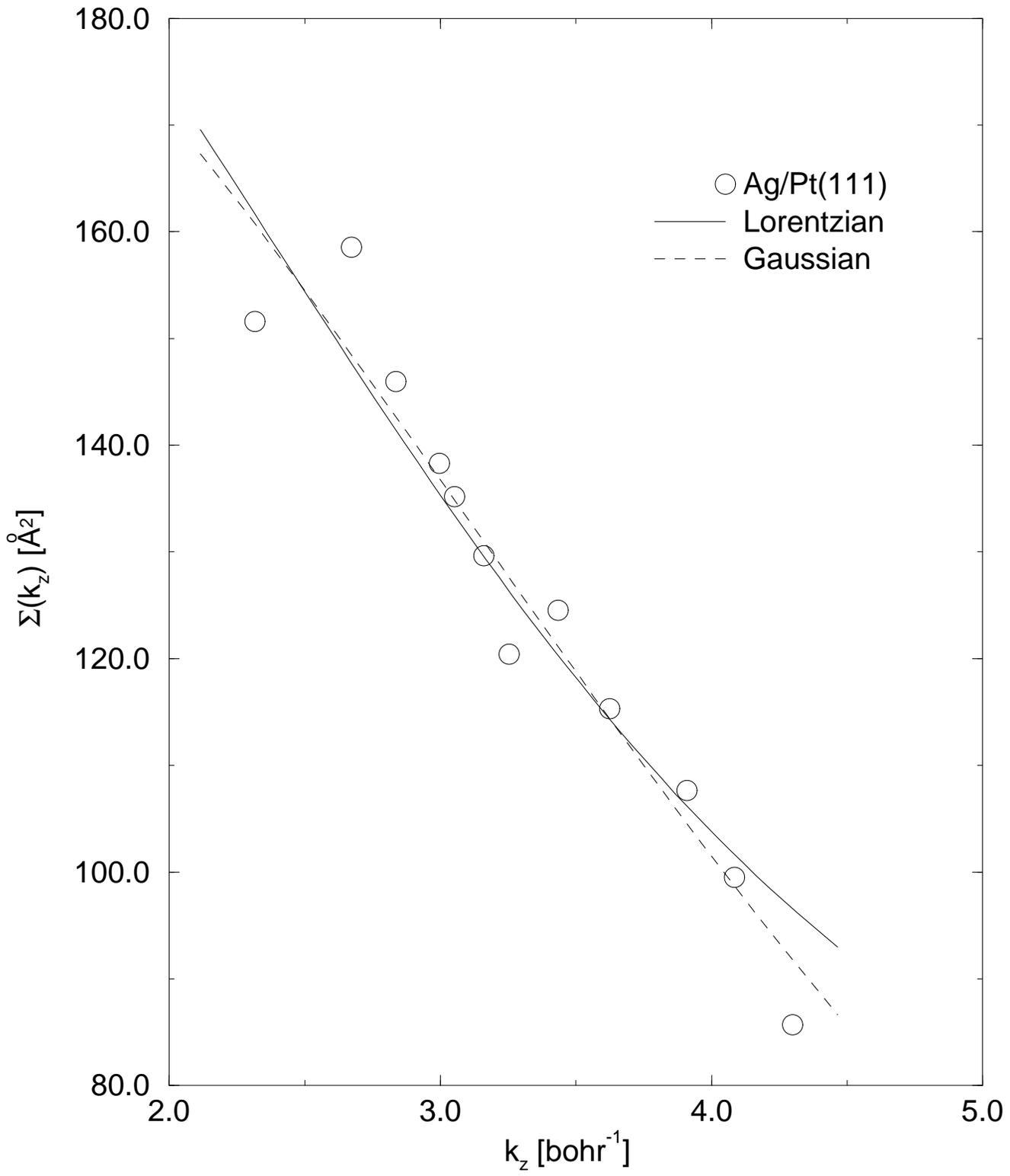
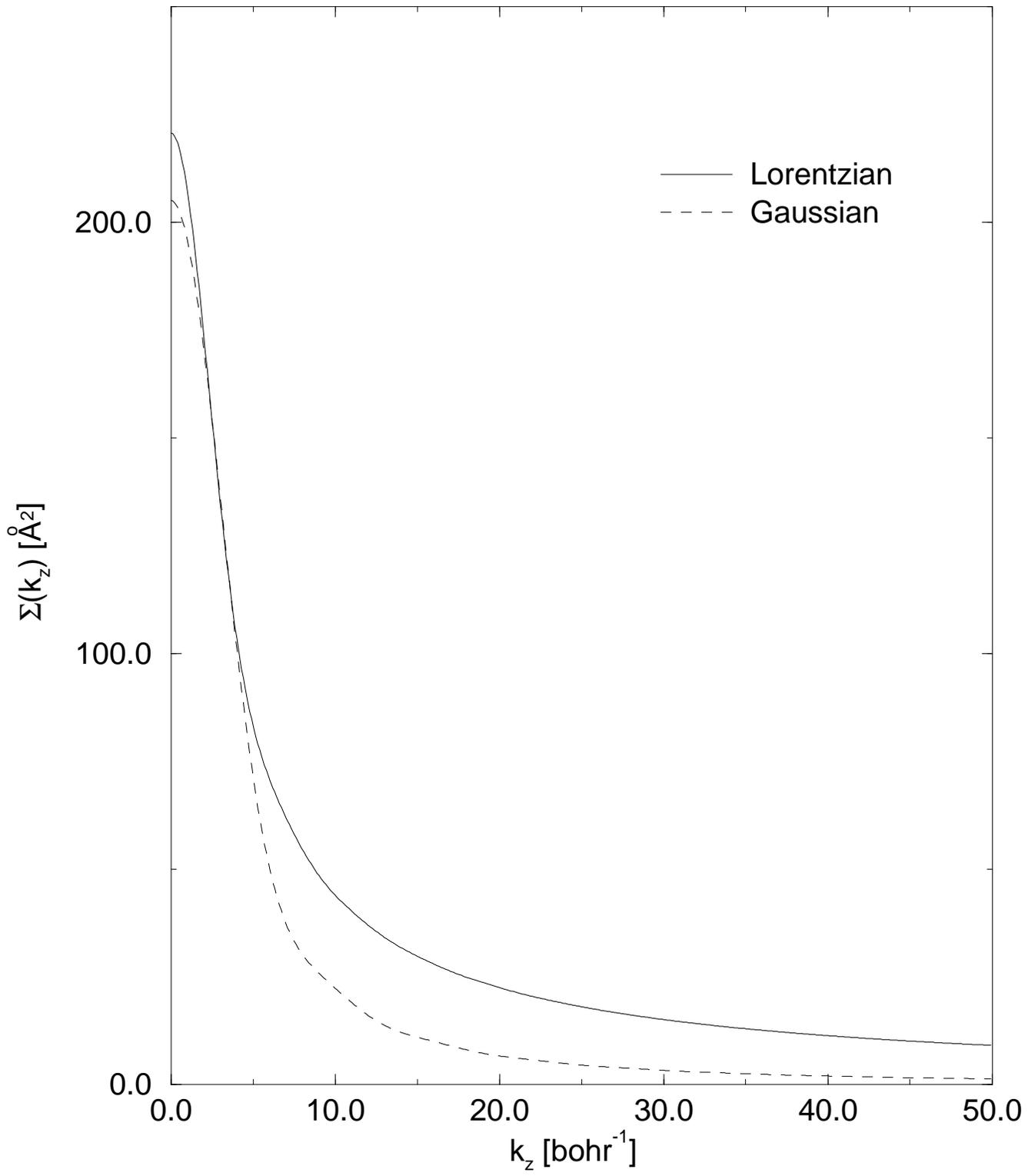
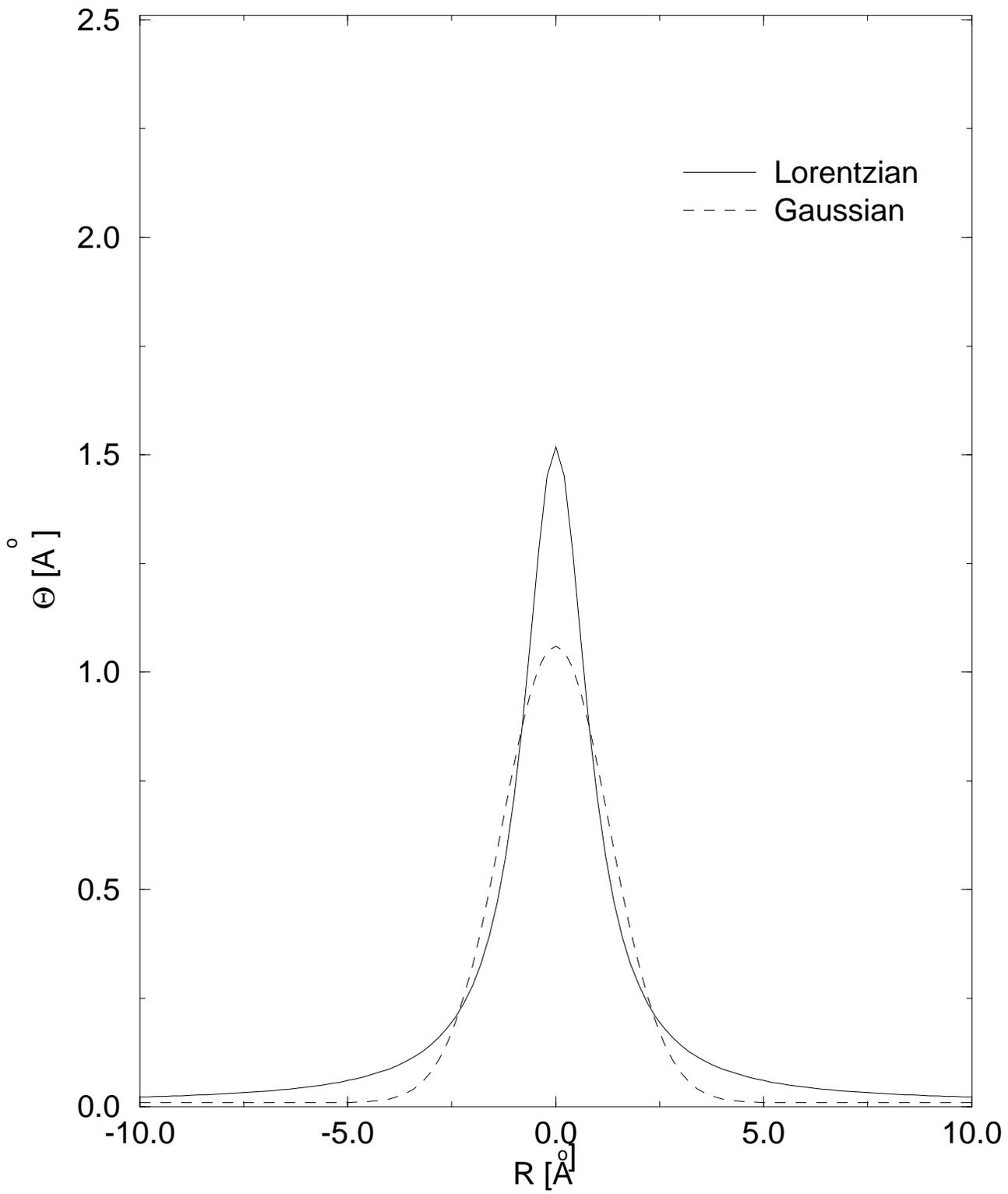
\end{document}